# SUBCRITICAL SUPERSTRINGS

W. Siegel[1]

*Institute for Theoretical Physics*
*State University of New York, Stony Brook, NY 11794-3840*

**ABSTRACT**

We introduce the Liouville mode into the Green-Schwarz superstring. Like massive supersymmetry without central charges, there is no $\kappa$ symmetry. However, the second-class constraints (and corresponding Wess-Zumino term) remain, and can be solved by (twisted) chiral superspace in dimensions D=4 and 6. The matter conformal anomaly is $c = 4-D < 1$. It thus can be canceled for physical dimensions by the usual Liouville methods, unlike the bosonic string (for which the consistency condition is $c = D \leq 1$).

---

[1]　Internet address: siegel@insti.physics.sunysb.edu.

# 1. MASSIVE SUPERPARTICLES

The basic idea behind relativistic mechanics is very simple: Just specify a set of constraints. Unlike nonrelativistic mechanics, the action contains no more information than this. For the usual massless superparticle [1] the constraints are $p^2 = 0$ and $d = 0$. The former constraint and the $\not{p}d$ half of the latter are first-class, the rest of the latter is second-class. First-class constraints generate gauge symmetries in the action: $p^2$ gives $\tau$-reparametrizations while $\not{p}d$ generates $\kappa$ symmetry. In terms of the wave function or field, the first-class constraints are the wave/field equations.

One way to generate mass from a massless theory is by dimensional reduction. In the case of supersymmetry, this introduces central charges, which preserve the size of the supersymmetry representation. Another way is to just add a mass term. This produces a different result for supersymmetry: It creates no central charges, and thus doubles the number of on-shell supersymmetry generators. This effect is reflected in the mechanics action by breaking $\kappa$ symmetry [2], which otherwise would have gauged away half of the components of $\theta$. In the field theory, it is the statement that $\not{p}d = 0$ is no longer a field equation, since $d\not{p}d = p^2 \neq 0$. (With a central charge $Z$, we can instead have $(\not{p} + mZ)d = 0$, and $d(\not{p} + mZ)d = p^2 + m^2Z^2 = 0$.)

In general, the superparticle constraints are difficult to solve covariantly [3,4]. In particular, the only practical way to solve the second-class constraints is to separate them into complex-conjugate halves, and solve half explicitly by eliminating half of the spinor coordinates, leaving a "chiral" superspace. (A more complicated possibility is to effectively reduce the components of $\theta$ by introducing an infinite number of ghost spinors of alternating statistics, which effectively add as $1 - 1 + 1 - ... = 1/2$.) For N=1 supersymmetry in ten dimensions (or eight Euclidean), this requires reducing the manifest Lorentz symmetry by two dimensions, so that the corresponding SO(2)=U(1) can be used to define complex $\theta$'s. However, for N=2 supersymmetry, or N=1 in D=4 or 6, the spinors are already complex. (In D=6 they are pseudoreal, and working in terms of complex spinors means breaking the internal SU(2), or USp(2N), symmetry.) This method is well known in 4D superfield theory, where chiral scalar superfields are the only fields with no gauge invariance: They are thus irreducible off-shell, or even on-shell in the massive case.

In these cases we write the massive superparticle Lagrangian as

$$L = [\dot{x} - i(\dot{\bar{\theta}}\gamma\theta + \dot{\theta}\gamma\bar{\theta})] \cdot p - \tfrac{1}{2}g(p^2 + m^2)$$

Transforming to the chiral representation by

$$x \to x - i\theta\gamma\bar{\theta}$$

we get

$$L = \dot{x} \cdot p - 2i\dot{\theta}\slashed{p}\bar{\theta} - \tfrac{1}{2}g(p^2 + m^2)$$

Since $\slashed{p}$ is now invertible (its determinant is a power of $p^2$, which is $-m^2$ on shell), we can make the invertible change of variables

$$\pi(\bar{\theta}) \equiv -2i\slashed{p}\bar{\theta}$$

to obtain the final form of the Lagrangian

$$L = \dot{x} \cdot p + \dot{\theta}\pi - \tfrac{1}{2}g(p^2 + m^2)$$

If the original $\theta, \bar{\theta}$ described an N=2 action, the final $\theta$ can be transformed into the real part of the original $\theta$, or simply reinterpreted as real, so the final action is then a real N=1 action. The final action has no second-class constraints, and only $p^2 + m^2 = 0$ as a first-class constraint.

Since the mechanics is completely described by the constraints, an even simpler analysis can be given directly in terms of them. What we have done is to solve the constraint $\bar{d} = 0$ for $\bar{\pi}$ in terms of $\theta$ (trivial), while solving $d = 0$ for $\bar{\theta}$ in terms of $\pi$ (requiring $p^2 \neq 0$), thus solving all spinor constraints, in terms of complex spinor variables, for their complex conjugates. The fact that $p^2 + m^2 = 0$ is the only remaining constraint then means that the result of quantization is the usual $1/(p^2 + m^2)$ propagator. The wave function (field) is a function of the remaining coordinates, $x$ and $\theta$ (and not $\bar{\theta}$).

To verify these results, we now give the equivalent derivation from the (free) field theory point of view. The simplest case is the 4D N=1 (or 3D N=2) massive chiral scalar multiplet. The usual free Lagrangian is

$$L = \int d^4\theta \ \bar{\phi}\phi + \int d^2\theta \ \tfrac{1}{2}m\phi^2 + \int d^2\bar{\theta} \ \tfrac{1}{2}m\bar{\phi}^2$$

The $\bar{\phi}$ field equation $d^2\phi = m\bar{\phi}$ can be solved for $\bar{\phi}$ in terms of $\phi$, thus treating $\bar{\phi}$ as auxiliary but $\phi$ as physical. The resulting Lagrangian is

$$L = -\tfrac{1}{m}\int d^2\theta \ \tfrac{1}{2}\phi(\Box - m^2)\phi$$

(This same method has been applied previously to spinors in electrodynamics [5].) As for the mechanics action, when interpreted as a 3D N=2 theory $\theta$ can now be taken as real (N=1). The same result can be derived by starting with this multiplet in terms of 3D N=1 superfields:

$$L = \int d^2\theta \, [\bar{\phi}d^2\phi + \tfrac{1}{2}m(\phi^2 + \bar{\phi}^2)]$$

becomes, after integrating out $\bar{\phi}$,

$$L = -\tfrac{1}{m}\int d^2\theta \, \tfrac{1}{2}\phi(\Box - m^2)\phi$$

where in this case $\theta$ is already real.

The next simplest example is the 4D N=2 (or 6D N=1) massive vector multiplet. The result was obtained in our previous paper [6] by N=1 superfield methods: There we chose a complex unitary gauge for a Higgs model of an irreducible massive N=2 vector multiplet described by a massless N=2 U(n) Yang-Mills multiplet coupled to n N=2 fundamental scalar multiplets. The resulting kinetic term (or the full Lagrangian for the corresponding Stueckelberg model) is

$$-\int d^4\theta \, \tfrac{1}{2}V(\Box - m^2)V$$

Although $V$ is treated as real (it was originally real, and no $\bar{V}$'s appear), the complex gauge condition requires that some external states be imaginary.

## 2. LIOUVILLE-GREEN-SCHWARZ STRINGS

The same classical mechanics analysis can be performed for strings. For superstrings, the requirement that the second-class constraints (or the supersymmetry generators) have the same algebra as for the superparticle requires D=3,4,6, or 10 already at the classical level [7], by the same Fierz identity that follows from closure of the first-class constraints for critical strings [8,3]. The existence of chiral spinors further restricts this to D=4 or 6. This means that only in D=4 or 6 can we divide the spinors in half in a way such that we can find second-class constraints satisfying $\{D, D\} = \{\bar{D}, \bar{D}\} = 0$, while $\{D, \bar{D}\} = \slashed{P}$.

We first transform to chiral superspace, and check that all spinor constraints can be solved. The Green-Schwarz action [8,3] in chiral superspace is, for N=1 in D=4 or 6,

$$L_{GS} = -\tfrac{1}{2}g^{mn}[\partial_m x - 2i(\partial_m\theta)\gamma\bar{\theta}] \cdot [\partial_n x - 2i(\partial_n\theta)\gamma\bar{\theta}] - 2i\epsilon^{mn}(\partial_m\theta)(\partial_n\slashed{x})\bar{\theta}$$

after performing the same shift on $x$ as above, and integrating derivatives off of $\bar{\theta}$ by parts in the Wess-Zumino term. ($g^{mn}$ has its determinant normalized to $-1$.) Now $\bar{\theta}$ appears nowhere with derivatives on it, corresponding to the constraint $\bar{\pi} = 0$. This Lagrangian can be rewritten as

$$L_{GS} = -\tfrac{1}{2}g^{mn}(\partial_m x) \cdot (\partial_n x) + 2i(g^{mn} - \epsilon^{mn})(\partial_m \theta)\gamma \cdot [(\partial_n x) - i(\partial_n \theta)\gamma\bar{\theta}]\bar{\theta}$$

$g^{mn} - \epsilon^{mn}$ projects out one chirality of the world-sheet derivative in $\partial_m \theta$ and the opposite in $\partial_n x + ....$ In the conformal gauge, or by simply using zweibeins, the second term is $4i(\partial_- \theta)(\partial_+ \slashed{x} - ...)\bar{\theta}$.

In the usual Green-Schwarz action $[\partial_+ x - 2i(\partial_+ \theta)\gamma\bar{\theta}]^2 = 0$ is one of the Virasoro constraints (from varying $g_{mn}$), but if the total Lagrangian has additional terms from the Liouville mode, then $[\partial_+ x - 2i(\partial_+ \theta)\gamma\bar{\theta}]^2$ equals minus the Liouville contribution to the corresponding part of the Virasoro operators. The statement that $[\partial_+ x - 2i(\partial_+ \theta)\gamma\bar{\theta}]^2$ is nonvanishing is equivalent to the statement that $\kappa$ symmetry is broken. Then we can write the $\partial x - i(\partial\theta)\gamma\bar{\theta}$ that appears in the $g^{mn} - \epsilon^{mn}$ term above as $[\partial x - 2i(\partial\theta)\gamma\bar{\theta}] + i(\partial\theta)\gamma\bar{\theta}$, and invert $\gamma \cdot [\partial x - 2i(\partial\theta)\gamma\bar{\theta}]$. So these terms can be rewritten as $(\partial_- \theta)\pi$, since $\bar{\theta}$ can be expressed locally as a (finite) polynomial in $\pi$.

The Green-Schwarz (non-Liouville) part of the action ($L = L_{GS} + L_{Liouville}$) then becomes

$$L_{GS} = -\tfrac{1}{2}g^{mn}(\partial_m x) \cdot (\partial_n x) + (\partial_- \theta)\pi$$

where for "$(\partial_- \theta)\pi$" we can use any of the various forms

$$(e_-{}^m \partial_m \theta)\pi, \quad (g^{mn} - \epsilon^{mn})(\partial_m \theta)\pi_n, \quad (g^{m0} - \epsilon^{m0})(\partial_m \theta)\pi$$

where the first form uses the zweibein, the second uses only the metric but uses a world-sheet vector $\pi$ (of which half is projected out, and so can be gauged away), and the last uses only the metric but is not manifestly world-sheet covariant (it is a gauge choice for the second form). The action is supersymmetric under the usual chiral-representation transformations: e.g., with the $\pi_m$ form,

$$\delta\theta = \zeta, \quad \delta\pi_m = (\partial_m \slashed{x})\bar{\zeta}, \quad \delta x = \theta\gamma\bar{\zeta}$$

As for the usual Green-Schwarz action, the Wess-Zumino term (now $\epsilon(\partial\theta)\pi$) is supersymmetric only after integration by parts.

The argument we have given above is for the Green-Schwarz part of the heterotic string Lagrangian (with $\theta$ of only one handedness), but generalizes directly for other superstrings. (Actually, for the heterotic case we should also include the usual chiral matter to treat the conformal anomaly in the nonsupersymmetric direction.) A simple way to collect all the $\theta$-(and $\bar{\theta}$-)dependent terms is to look at the variation of the action, which is manifestly covariant (although the Wess-Zumino term of the action itself is not), and integrate back. In the chiral representation, the covariant variation of $x$ is $\delta x - 2i[(\delta\theta_L)\gamma\bar{\theta}_L + (\delta\theta_R)\gamma\bar{\theta}_R]$ (with no $\delta\bar{\theta}$'s), while the $\theta$ and $\bar{\theta}$ variations get no covariantizations, and all $\theta$ terms in the action have $\bar{\theta}$'s, so all terms in the action except $-\frac{1}{2}g(\partial x)^2$ can be obtained by integrating the $\delta\bar{\theta}$ terms. Explicitly, these terms are

$$iJ_m \cdot [(g^{mn} - \epsilon^{mn})(\partial_n\theta_L)\gamma\delta\bar{\theta}_L + (g^{mn} + \epsilon^{mn})(\partial_n\theta_R)\gamma\delta\bar{\theta}_R],$$

$$J \equiv \partial x - 2i[(\partial\theta_L)\gamma\bar{\theta}_L + (\partial\theta_R)\gamma\bar{\theta}_R]$$

so the Green-Schwarz part of the action can be written as

$$L_{GS} = -\tfrac{1}{2}g^{mn}(\partial_m x)\cdot(\partial_n x) + i\hat{J}_m \cdot [(g^{mn} - \epsilon^{mn})(\partial_n\theta_L)\gamma\bar{\theta}_L + (g^{mn} + \epsilon^{mn})(\partial_n\theta_R)\gamma\bar{\theta}_R],$$

$$\hat{J} \equiv \partial x - i[(\partial\theta_L)\gamma\bar{\theta}_L + (\partial\theta_R)\gamma\bar{\theta}_R]$$

Again, $\hat{J}$ differs from the covariant $J$, which appears in the matter part of the Virasoro generators $J_\pm^2$, by $\bar{\theta}$ terms. As before, $J_\pm$ can be inverted, allowing redefinitions that give the final form of the Green-Schwarz part of the action

$$L_{GS} = -\tfrac{1}{2}g^{mn}(\partial_m x)\cdot(\partial_n x) + (\partial_-\theta_L)\pi_L + (\partial_+\theta_R)\pi_R$$

The roles of $\theta$ and $\bar{\theta}$ can be reversed for either the left or right handednesses (with corresponding sign changes in the $\theta\gamma\bar{\theta}$ shifts in $x$), so by switching just one handedness we can obtain a "twisted" chiral representation.

By the usual analysis of quadratic Virasoro operators, and noting that the number of components of a chiral spinor is half the number of the usual $2(D-2)$-component spinor of the Green-Schwarz action, we find that the "matter" ($x$ and $\theta$) part of the coefficient of the conformal anomaly is

$$c = D - 2(D-2) = 4 - D$$

The intercept of the leading Regge trajectory also follows as in the usual analysis, with $c$ acting as the effective dimension:

$$\alpha_0 = \frac{c-1}{24} = \frac{3-D}{24}$$

(This is the intercept of the Regge trajectory for the superspin: The ground state is superspin 0, which contains spins 0 and higher.) We thus have Liouville theories for D = 4 or 6, with $c = 0$ or $-2$, and $\alpha_0 = -1/24$ or $-1/8$. This is consistent with the usual quantum condition $c \leq 1$ found for Liouville theories in the continuum [9] and lattice [10] approaches, and the no-tachyon condition $\alpha_0 \leq 0$ is consistent with supersymmetry. (The ground-state mass$^2$ is proportional to $-\alpha_0$. If we could extend this analysis to D=3 we would find $c$=1 and $\alpha_0$=0, giving a massless ground state. This might be possible in a light-cone analysis, since the massless 3D N=1 scalar multiplet, unlike general multiplets with more supersymmetries and no central charges, has the same number of states as the massive multiplet.)

The $g^{mn}$ terms of the matter (Green-Schwarz) sector determine the field content and propagator of the corresponding random matrix model. If we look at the type II superstring in twisted chiral superspace for D=4, we find the same field content and propagator as in the 4D random supermatrix model described previously [6]. The lowest order terms for the Lagrangian of the supermatrix model found there were

$$-\tfrac{1}{2}V(\Box - m^2)V + [-\tfrac{1}{2}m^2 V^3 + (\bar{d}^{\dot\alpha} V)V i \partial_{\alpha\dot\alpha} d^\alpha V]$$

where $V$ is a "real" N=1 (or twisted-chiral N=2) superfield, as we discussed for the massive N=2 superparticle. However, to find the derivative part of the cubic terms we still need to understand how to handle the Wess-Zumino terms ($\epsilon(\partial\theta)\pi$) of the string action [11]. We may also need to consider the fact that vertex operators for 10D light-cone Green-Schwarz formalism (such as appear in the string field theory) do not appear straightforwardly, probably because of subtleties in the quantization. Another possibility is that the conformal anomaly that generates the Liouville mode is accompanied by a $\kappa$ anomaly that modifies the Wess-Zumino term. Also, although our model has no $\kappa$ symmetry, the world-sheet chirality of the second-class constraints constrains the form of the string action (it specifies the Wess-Zumino term), and so may require additional terms quantum mechanically. Since the derivative term in the random matrix action is required by unitarity, as follows from quantization of the gauge field theory, we expect that a more careful treatment of the quantization of the string mechanics action would also require such a term. Thus, the identification of the world-sheet metric terms in the string action with the kinetic and nonderivative interaction terms in the random matrix model, together with unitarity of the matrix model, may imply that the full string theory and matrix model are equivalent.

Bilal and Gervais [12] claimed to have found an operator formalism for a D=5 subcritical superstring, but their supersymmetry transformations are six-dimensional, treating the Liouville mode as one of the coordinates. However, the action is not invariant under translation of this mode (even in the absence of a cosmological term), which generates scaling of the string coupling constant, and therefore not under this supersymmetry. Also, our superstrings differ from theirs in that we have $c \leq 1$, which is the range consistent with the usual Lagrangian quantization of the Liouville theory. Kutasov and Seiberg [13] also discussed noncritical superstrings, but they used a Ramond-Neveu-Schwarz (RNS) formulation where the spacetime interpretation is not clear. In particular, the condition D=3,4,6, or 10 (which is required for closure of the supersymmetry algebra even in the RNS approach) is not evident in their treatment. Also, the RNS formalism is not suitable for random-lattice quantization because of the difficulty of treating world-sheet fermions on a lattice.

## ACKNOWLEDGMENTS

I thank Nathan Berkovits, Jan de Boer, and Dileep Jatkar for discussions. This work was supported in part by the National Science Foundation Grant No. PHY 9309888.